\documentclass[aps,showpacs,amsmath,amssymb,11pt]{revtex4}

\usepackage{latexsym} % Gets \Box etc
\usepackage{amssymb}  % \gtrsim, \geqslant, etc etc: see amsguide.ps
\usepackage{amsbsy}   % \pmb and \boldsymbol
\usepackage{graphicx}       % For PostScript figures
\usepackage[dvips]{color}
\usepackage{bm}   % Bold math: \bm{1}
\newcommand\eqn[1]{(\ref{#1})}      % parentheses around the LaTex "ref" macro
\newcommand{\e}{{\rm e}}   % 2.718281828

\newcommand{\be}{\begin{equation}}
\newcommand{\ee}{\end{equation}\noindent}
\newcommand{\bea}{\begin{eqnarray}}
\newcommand{\eea}{\end{eqnarray}\noindent}

\newcommand{\E}{{\mathcal E}}
\newcommand{\tr}{\mbox{tr}}

\newcommand{\nn}{\nonumber \\}

\makeatletter %\catcode`\@=11

\def\appendix{\par                              % Have \appendix say
    \setcounter{section}{0}                     % `Appendix A', not just `A'
    \setcounter{subsection}{0}
    \renewcommand{\theequation}{\Alph{section}.\arabic{equation}}
    \renewcommand{\thesection}{Appendix \Alph{section}
                \setcounter{equation}{0}  } %Have eqns numbered (A.1) etc
    \renewcommand{\thesubsection}{\Alph{section}.\arabic{subsection}}
}

\def\applabel#1{\@bsphack
  \protected@write\@auxout{}%
         {\string\newlabel{#1}{{\Alph{section}}{\thepage}}}%
  \@esphack}
\def\subsection{\@startsection{subsection}{2}{\z@}{-3.25ex plus -1ex minus
 -.2ex}{1.5ex plus .2ex}{\normalsize\bf}}

\def\subsubsection{\@startsection{subsubsection}{3}{\z@}{-3.25ex plus
 -1ex minus -.2ex}{1.5ex plus .2ex}{\normalsize}}

\makeatother   %\catcode`\@=12

\begin{document}

\title{\bf Gutzwiller's Trace Formula and Vacuum Pair Production}

\author{Dennis D.\ Dietrich and Gerald V.\ Dunne}
\affiliation{Institut f\"ur Theoretische Physik, Universit\"at Heidelberg, 
Philosophenweg, 69120 Heidelberg, Germany}

%\date{June 11, 2007}
\date{\today}

\begin{abstract}
We propose a new application of the Gutzwiller trace formula formalism, to give a compact expression for the semiclassical  vacuum pair production rate in quantum electrodynamics,
for general inhomogeneous electromagnetic background fields.
%using the formalism of the Gutzwiller trace formula. 
%The full prefactor is naturally expressed in terms of a certain monodromy matrix.
\end{abstract}

%\begin{titlepage}
\maketitle
%\renewcommand{\thepage}{}          % No page number on title page

%\end{titlepage}

The Gutzwiller trace formula has found a wide variety of applications in theoretical and mathematical physics \cite{gutzwiller,littlejohn,cvitanovic}. Here we point out a new area where its language provides insight and simplification to a difficult computational problem in relativistic quantum field theory.
Vacuum polarization effects in quantum electrodynamics (QED) predict that electron-positron pairs can be produced from the vacuum in the presence of a classical electric field. This remarkable phenomenon was predicted, and its rate estimated, for uniform fields in \cite{he,schwinger,gvd}, but has not been directly observed as the rate is tiny for accessible field strengths. It is conceivable that sufficiently strong electric fields may be reached in X-Ray Free Electron Lasers
\cite{ringwald}, but such fields have strong temporal and spatial inhomogeneities. Unfortunately, very little is known about the rate when the  background field has such general inhomogeneities. On the other hand,  in the approximation where the background electric field has a fixed direction and a magnitude that varies in just one dimension, either spatial or temporal, one can use WKB-based  techniques \cite{nikishov,brezin,kimpage}. A promising approach for going beyond this one-dimensional case is the "worldline instanton" method \cite{affleck,wli}, based on an instanton approximation to Feynman's worldline path integral formulation of QED \cite{feynman}.
Another related approach is a direct Monte Carlo evaluation of the
worldline form of the effective action \cite{giesklingmuller}. 
In this note we propose a new approach to this problem, based on a close connection between  the worldline instanton approach and the Gutzwiller trace formula \cite{gutzwiller,littlejohn,cvitanovic}. This connection gives a well-defined computational strategy for treating multi-dimensional inhomogeneities in the background electromagnetic field.

The technical 
problem is to compute the imaginary part of the effective action in the classical electromagnetic background field, from which the vacuum pair production rate follows  \cite{schwinger}: $P_{\rm production}=1-e^{-2\, {\rm  Im}\, \Gamma}\approx 2\, {\rm  Im}\, \Gamma$.
%\be
%P_{\rm production}=1-e^{-2\, {\rm  Im}\, \Gamma}\approx 2\, {\rm  Im}\, \Gamma\quad.
%\label{decay}
%\ee
For example, for a constant electric field of magnitude ${\mathcal E}$, the leading weak field result (we consider scalar QED) is \cite{he,schwinger,gvd}
\be
 \frac{{\rm  Im}\, \Gamma}{\rm Vol} \sim \frac{e^2\,{\E}^2}{16\pi^3} \, e^{-\frac{m^2\pi}{e \E}}\quad .
 \label{constant}
 \ee
 The basis of our proposal is the worldline formalism of QED \cite{feynman,halpern,csreview}, in which the effective action is expressed in terms of a
{\it quantum mechanical} path integral in four-dimensional Euclidean space, with paths $x_\mu(\tau)$ parametrized by proper-time $\tau$. This approach has led to many beautiful advances in our understanding of perturbative scattering amplitudes \cite{csreview}, but here we propose to use it to extract non-perturbative information.
The  effective action for a scalar charged particle (charge $e$, mass $m$) in a Euclidean classical gauge background $A_\mu(x)$ is the functional ($D_\mu=\partial_\mu + i e A_\mu$ is the covariant derivative): 
\bea
\label{eff}
\Gamma [A] &=&-\tr \, \ln \left(-D_\mu^2+m^2\right)\\
&=& \int_0^{\infty}\frac{dT}{T}\, \e^{-m^2T}\int d^4x^{(0)} \, \langle x^{(0)}| e^{-T(-D_\mu^2)}\, |x^{(0)}\rangle \nn
&=&
\int_0^{\infty}\frac{dT}{T}\, \e^{-m^2T}\int d^4x^{(0)}
\!\!\!\!\!\!\! \int\limits_{x(T)=x(0)=x^{(0)}}\!\!\!\!\!\!\!\!\!\! {\mathcal D}x
\,\, {\rm exp}\left[-\int_0^Td\tau
\left(\frac{\dot x^2_\mu}{4} + e\, A_\mu \dot x_\mu \right)\right]
\nonumber
\eea
In the last line, the trace of the associated Euclidean propagation operator has been written as a functional integral $\int {\mathcal D}x$ over all closed Euclidean spacetime paths $x_\mu(\tau)$ that are periodic (with period $T$) in the proper-time parameter $\tau$ \cite{feynman}.  We use the QED worldline path integral normalization conventions of \cite{csreview}. 

The strategy of the worldline instanton method \cite{wli} is to evaluate the quantum mechanical path integral in \eqn{eff}
semiclassically \cite{levit}, and then to evaluate each of the $T$ and $x^{(0)}$ integrals by steepest descents. These are precisely the steps  in deriving the Gutzwiller trace formula \cite{gutzwiller,littlejohn,cvitanovic}, although there one is concerned with a non-relativistic  Schr\"odinger operator rather than the Euclidean Klein-Gordon operator, an oscillatory amplitude $e^{i\, S/\hbar}$ rather than the Euclidean form $e^{-S}$, and the trace of the resolvent rather than the trace of the logarithm. Nevertheless, despite these differences, in this note we show that the worldline instanton computation can usefully be formulated in the language of the Gutzwiller trace formula.

The first step is to make a semiclassical approximation for the propagation kernel
\bea
K(x, x^\prime; T):=\langle x| e^{-T(-D_\mu^2)}\, |x^\prime\rangle
\approx \frac{1}{(2 \pi)^2}\,
\sqrt{\left| \det\left(\frac{\partial^2 R}{\partial x\, \partial x^\prime}\right)\right |}\, e^{-R(x, x^\prime; T)}\quad ,
\label{semi}
\eea
where $R(x, x^\prime; T)$ is the Hamilton principal function for the classical trajectory  from $x$ to $x^\prime$ in four-dimensional Euclidean
space, in the proper-time interval $T$. This classical trajectory is obtained by solving the Euclidean classical equations of motion
\bea
\ddot{x}_\mu=2e F_{\mu\nu}(x)\, \dot{x}_\nu \quad, \quad (\mu,\nu=1\dots4)\quad,
\label{euler}
\eea
where $F_{\mu\nu}=\partial_\mu A_\nu-\partial_\nu A_\mu$ is the background field strength. To evaluate the trace in \eqn{eff} we will need the diagonal propagation kernel $K(x^{(0)}, x^{(0)}; T)$,
but for now we consider the point-split propagation from $x$ to $x^\prime$. The classical equations of motion \eqn{euler} are those for a charged particle moving in an inhomogeneous electromagnetic field $F_{\mu\nu}(x)$, so the "energy" is conserved on a classical trajectory : $E=\frac{1}{4}\,\dot{x}_\mu^2 = {\rm constant}$.
%\bea
%E=\frac{1}{4}\,\dot{x}_\mu^2 = {\rm constant}
%\label{energy}
%\eea

The next step is to perform the $T$ integral by steepest descents. The critical point of the exponential factor arises when $\frac{\partial R}{\partial T}=-m^2$. This has a natural classical interpretation in terms of the Legendre transformation between the Hamilton principal function $R(x, x^\prime; T)$ [expressed in terms of the total time elapsed along the trajectory] and the action $S(x, x^\prime; E)$ [expressed in terms of the constant energy of the trajectory] : $R(x, x^\prime; T)=S(x, x^\prime; E)-E\, T$.
It follows that $\frac{\partial R}{\partial T}=- E$, and $ \frac{\partial S}{\partial E}=T$.
%\bea
%\frac{\partial R}{\partial T}=- E\quad , \quad \frac{\partial S}{\partial E}=T \quad .
%\label{var}
%\eea
Thus,  the critical point $T_c$ of the $T$ integral occurs when $E=m^2$, so that 
\bea
\int_0^\infty \frac{dT}{T}e^{-m^2 T} K(x, x^\prime; T)\approx 
\frac{1}{(2 \pi)^2\, T_c}\, \sqrt{\left| \det\left(\frac{\partial^2 R}{\partial x\, \partial x^\prime}\right)\right
|_{T_c}}\, \sqrt{\frac{2\pi}{\left | \frac{\partial^2 R}{\partial T^2}\right |_{T_c}}}\, e^{-S(x, x^\prime; m^2)}
\quad ,
\label{tint}
\eea
up to a possible phase that we discuss later. The two prefactor contributions combine in a simple way if we consider coordinates $x^{(0)}_\parallel$ along the classical trajectory, and  $x^{(0)}_\perp$ transverse to the trajectory. Then \cite{gutzwiller,littlejohn,cvitanovic}
\bea
\left . \frac{\det\left(\frac{\partial^2 R}{\partial x\, \partial x^\prime}\right)}{\frac{\partial^2 R}{\partial T^2}}\right |_{T_c} =\frac{1}{\dot{x}_\parallel\, \dot{x}_\parallel^\prime}\, \det\left(\frac{\partial^2 S(x, x^\prime; m^2)}{\partial x_\perp\, \partial
x_\perp^\prime}\right)\quad .
\label{combin}
\eea

The final step is the coincident limit $x\to x^\prime =x^{(0)}$, and trace over $x^{(0)}$. This trace is
also done by steepest descents and implies that the closed loop is in fact periodic
\cite{gutzwiller,littlejohn,cvitanovic}. Periodic solutions to \eqn{euler} are known as {\it worldline instantons} \cite{wli}. From \eqn{combin}, the integration over $x_\parallel^{(0)}$ yields a factor $\int dx^{(0)}_\parallel/\dot{x}^{(0)}_\parallel=T_c/2$
[reparametrization invariance of the periodic orbit], while the $x_\perp^{(0)}$ integral produces another determinant factor. Remarkably, this determinant factor combines with the remaining transversal determinant factor in \eqn{combin} to give:
\bea
 \frac{\det\left(\frac{
\partial^2 S(x, x^\prime; m^2)}{\partial x_\perp\, \partial x_\perp}+
\frac{\partial^2 S(x, x^\prime; m^2)}{\partial x_\perp^\prime\, \partial x_\perp}+
\frac{\partial^2 S(x, x^\prime; m^2)}{\partial x_\perp\, \partial x_\perp^\prime}+
\frac{\partial^2 S(x, x^\prime; m^2)}{\partial x_\perp^\prime\, \partial
x_\perp^\prime}\right)}{\det\left(\frac{\partial^2 S(x, x^\prime; m^2)}{\partial x_\perp\, \partial
x_\perp^\prime}\right)}&=& 
\det\left(\frac{\partial\left(p_\perp-p_\perp^\prime, x_\perp-x^\prime_\perp\right)}{\partial\left(x_\perp^\prime, p_\perp^\prime\right)}\right) \nn
&=:& \det\left({\bf 1}-J\right)\quad ,
\eea
where all determinants are evaluated at vanishing transverse displacements. Here $J$ is the {\it monodromy matrix}, for a $6$-dimensional surface of section in phase space transverse to the periodic phase space  orbit with constant energy $E=m^2$. Consider an initial transverse displacement $\left(\begin{matrix} \delta x_\perp^\prime \\ \delta p_\perp^\prime\end{matrix}\right)$ from a point on the closed orbit in phase space, and evolve for time $T$, and the final displacement from the orbit is related to the initial one by the monodromy matrix: $\left(\begin{matrix} \delta x_\perp^{\prime\prime} \\ \delta p_\perp^{\prime\prime}\end{matrix}\right)= J \left(\begin{matrix} \delta x_\perp^\prime \\ \delta p_\perp^\prime\end{matrix}\right)$. Putting all these parts together, and collecting phases carefully \cite{wli}, one obtains a compact final expression :
 \bea
{\rm Im} \, \Gamma\approx \frac{e^{-S(E=m^2)}}{\sqrt{\det\left({\bf
1}-J\right)}}\quad . 
\label{final}
\eea
The principal advantage of expressing the computation in this language of the Gutzwiller trace formula  is that the total prefactor is encapsulated in a {\it single determinant}, which moreover has a natural mathematical and geometrical meaning
in the Euclidean phase space. In previous work \cite{brezin,kimpage,wli}  the various prefactor contributions
have been evaluated separately, and then combined at the end. Thus, the computational strategy is as follows:
\begin{enumerate}
\item solve the classical equations of motion in four dimensional Euclidean space to
find all closed periodic trajectories of energy $E=m^2$ : the ``worldline instanton(s)''.
\item evaluate the classical action $S(E=m^2)$ on these trajectories. The dominant contribution comes from the trajectory(ies) with largest $e^{-S(m^2)}$.
\item compute the prefactor from the monodromy matrix $J$ for the dominant trajectory(ies).
\end{enumerate}

The only concrete comparison we can make is to compute ${\rm Im} \, \Gamma$ for the case of a one-dimensional inhomogeneity, which can be computed in several other ways \cite{brezin,kimpage,wli}.  Consider, for example, the case of a time dependent electric field directed in the $x_3$ direction.
We can choose a Euclidean gauge field
$
A_3(x_4)=\frac{\E}{\omega}\, f(\omega~x_4)
$,
where $\E$ characterizes the overall magnitude of the associated electric field, $\omega$ characterizes the scale of the time dependence, and $f(\omega\, x_4)$ is some smooth function. For example, for a constant electric field $\E(t)=\E$, we have $f(x)=x$; for a sinusoidal electric field $\E(t)=\E\,
\cos(\omega~t)$, we have $f(x)=\sinh(x)$; and for a single-pulse electric field $\E(t)=\E\, {\rm
sech}^2(\omega~t)$, we have $f(x)=\tan(x)$. Then the classical action on a periodic trajectory of energy $E$ can be written [here $y:=\frac{e \E}{\omega \sqrt{E}}\, f(x)$]
\bea
S(E)=\oint d x_4 \sqrt{E-\left(\frac{e \E}{\omega}\, f(\omega \, x_4)\right)^2}=\frac{2E}{e\, \E}\int_{-1}^1 dy\,
\frac{\sqrt{1-y^2}}{f^\prime(x(y))}
\eea
This is precisely the exponent appearing in the standard result for the pair
production rate \cite{brezin,kimpage,wli}.
To evaluate the prefactor, we can choose $x_4$ as $x_\parallel$. Then the transverse $x_3$ direction is in fact an invariant "flat" direction, so we do not need to perform the transverse integration. 
This illustrates  the important point that \eqn{final} must be interpreted appropriately when there are physical zero modes.
Thus, we go back to 
\eqn{tint} and observe that $\frac{\partial^2 R}{\partial T^2}=-1/\frac{\partial^2 S}{\partial E^2}$. Furthermore, the other determinant factor in \eqn{tint}
is easily computed (see \cite{wli}b) using the Gel'fand-Yaglom formula:
\bea
\left . \det\left(\frac{\partial^2 R}{\partial x\, \partial x^\prime}\right)\right |_{x=x^\prime}=
\frac{m^4}{16\, E^3 \, T^2}\,
\frac{1}{\dot{x}_4^2 \left(\frac{\partial^2 S}{\partial E^2}\right)^2}\quad .
\eea
Thus, relative to the constant spatial volume $V_3$,
\bea
\frac{{\rm Im} \, \Gamma}{V_3}\approx \frac{\sqrt{2\pi}}{2(4\pi)^2 m} \left[ \frac{e^{-S(E)}}{\frac{\partial S}{\partial E}\sqrt{\frac{\partial^2 S}{\partial E^2}}} \right]_{E=m^2}\quad ,
\label{1dfinal}
\eea
%where the extra $T_c=\frac{\partial S}{\partial E}|_{E=m^2}$ in the denominator comes from
%the free ($x_1$, $x_2$) directions.
Note that (\ref{1dfinal}) agrees precisely with the conventional WKB result \cite{brezin,kimpage,wli}.

We now turn to a multi-dimensional example. Consider the two-dimensional
Euclidean problem (\ref{euler}) in the $x_3$-$x_4$ plane, with $F_{34}\equiv
F(r)$, where $r:=\sqrt{x_3^2+x_4^2}$. 
%The $x_1$ and $x_2$ directions  correspond to free motion. 
The associated Minkowski electric
field points along the $x_3$ axis and is a function of $\sqrt{t^2-x_3^2}$,
i.e. a configuration studied, e.g., in \cite{cooper}. There can
exist circular orbits centered around $r=0$. The fluctuation determinant for
fluctuations around such an orbit of radius $r_0$ follows from the
corresponding monodromy matrix $J$. In polar coordinates the circular orbit 
is characterized by $r(\tau)=r_0$, and 
$\dot\theta(\tau)\equiv\dot\theta_0=2F_0:=2F(r_0)$. Linearizing the equations 
of motion in fluctuations $\rho$ and $\vartheta$ around the periodic 
trajectory, where $r(\tau)=:r_0+\rho(\tau)$ and 
$\theta(\tau)=:\theta_0(\tau)+\vartheta(\tau)$, and solving the resulting 
equations for the intial conditions 
$\rho(0)=\delta x_\perp^\prime$ and $\dot\rho(0)=\delta p_\perp^\prime$, 
leads to the following solution for the radial fluctuations:
\be
\rho
=
\delta x_\perp^\prime\cos(2\tau F_0\sigma)
+
\delta p_\perp^\prime(2F_0\sigma)^{-1}\sin(2\tau F_0\sigma)\quad.
\label{deltar}
\ee
Here $\sigma:=[1+\frac{r_0}{F_0}(\partial_r|_{r_0}F)]^{1/2}$, and we 
made use of $\rho\dot\theta_0=\dot{\vartheta}r_0$, 
which follows from the conservation of the magnitude of the velocity 
$\sqrt{\dot x_3^2+\dot x_4^2}$. To compute the transverse deviation from the 
orbit after one cycle, in principle, we have to calculate the time needed in 
order to return to the same longitudinal coordinate ($\delta x_\parallel=0$), 
which here means $\theta(T+\delta T):=2\pi$, where $T=\pi/F_0$ is the period 
of the unperturbed orbit. Putting into Eq.~(\ref{deltar}) $\tau=T+\delta T$, 
instead of $\tau=T$, however, leads merely to corrections quadratic in the 
initial fluctuation parameters $\delta x_\perp^\prime$ and 
$\delta p_\perp^\prime$. Therefore, 
$\delta q_\perp^{\prime\prime}=\rho(T)$, and 
$\delta p_\perp^{\prime\prime}=\dot\rho(T)$. Then the monodromy matrix is:
\be
J=\left(\begin{array}{cc}
\cos(2\pi\sigma)&2F_0\sigma\sin(2\pi\sigma)\\
-(2F_0\sigma)^{-1}\sin(2\pi\sigma)&\cos(2\pi\sigma)\end{array}\right)\quad.
\ee
The corresponding fluctuation determinant is given by 
$\det(1-J)=4\sin^2(\pi\sigma)$. 

We conclude with some comments and open problems. (i) Finding closed 
periodic orbits to \eqn{euler} is non-trivial, but recasting the problem in 
phase space proves helpful. (ii) If the physical electric field is too 
localized in space, then we know physically that the pair production rate 
vanishes [since the virtual vacuum dipole pairs cannot gain enough energy 
from the field to become real electron-positron pairs]. In simple cases this 
corresponds to the non-existence of periodic classical Euclidean trajectories \cite{wli}. It 
would be interesting if this were more generally true: that the mere 
existence of such worldline instanton loops might be used as an {\it indicator} of pair 
production. (iii) The phases arising from the steepest descent integrals 
combine to give ${\rm Im}\,\Gamma$ in the cases where the electric field is a 
function of either $t$ or $\vec{x}$ \cite{wli}, but the general mixed case 
needs further analysis. (iv) If the gauge
field corresponding to the external field can be put into the non-linear 
gauge where $A_\mu^2(x)={\rm constant}(\equiv E)$, then we can solve the simpler
first-order equations $\dot{x}_\mu=-2eA_\mu(x)$, as was observed long ago by Nambu \cite{nambu}. 
(v) It would be interesting to extend our method to
inhomogeneous non-abelian fields, for which little 
is known beyond simple 
quasi-abelian cases. This suggests studying  the Wong equations
\cite{wong} describing the classical motion of a color-charged particle in a 
non-abelian background, which is a much richer mathematical system.

\vskip .5cm

{\bf Acknowledgements:} We gratefully acknowledge discussions with 
H.~Gies, Q.-h.~Wang, B.~Tekin, M.~Lohe and \"O.~Sar{\i}o\u{g}lu. GD thanks the DOE for support through the grant DE-FG02-92ER40716, the DFG for support through the Mercator Guest Professor Program, and the ITP at Heidelberg for hospitality.

\end{document}